# Design and optimization of tapered structure of near-field fiber probe based on FDTD simulation

H. NAKAMURA and T. SATO

Theory and Computer Simulation Center, National Institute for Fusion Science, 322-6 Oroshi-cho, Toki, Gifu 509-5292, Japan

H. KAMBE and K. SAWADA

Department of Applied Physics, Shinshu University, 500 Wakasato, Nagano 380-8553, Japan

T. SAIKI

Kanagawa Academy of Science and Technology, 3-2-1 Sakado, Takatsu, Kawasaki, Kanagawa 213-0012, Japan

**Summary**

The finite-difference time-domain method was employed to simulate light propagation in tapered near-field fiber probes with small metal aperture. By conducting large-volume simulations, including tapered metal-cladding waveguide and connected optical fiber waveguide, we illustrated the coupling between these guiding modes as well as the electric field distribution in the vicinity of the aperture. The high collection efficiency of a double-tapered probe was reproduced and was ascribed to the shortening of the cutoff region and the efficient coupling to the guiding mode of the optical fiber. The dependence of the efficiency on the tapered structure parameters was also examined.

## 1. Introduction

Improvement of the optical throughput and collection efficiency of aperture probes is the most important issue to be addressed for the application of near-field scanning optical microscopy (NSOM) in optical recording, fabrication, and manipulation as well as spectroscopic studies. The tapered region of the aperture fiber probe is considered to be the metal-cladding optical waveguide, whose propagation properties are characterized by the cutoff diameter and the absorption coefficient of the cladding metal. Through systematic experimental studies, it has been confirmed that the transmission efficiency decreases in the region where the core diameter is smaller than the wavelength of the propagating light. On the basis of this finding, we proposed to shorten the narrow metal-cladding region with strong optical losses by making a double-tapered structure with a large cone angle. This structure is easily realized using a multi-step chemical etching technique. It has been demonstrated that the transmission efficiency is much improved by 1-2 orders of magnitude as compared to the single-tapered probe with a small cone angle (Saiki *et al.*, 1996).

Further optimization of the tapered structure is needed to achieve much higher probe efficiency. However, it is very time-consuming to assess many structure parameters, such as the cone angle and taper length, by trial and error. Numerical analysis is a more reasonable way to attain an optimized structure efficiently and to understand the electromagnetic field distribution in a tapered waveguide including the vicinity of the aperture. Computational calculation by the finite-difference time-domain (FDTD) method is the most popular and promising method available for this purpose (Furukawa & Kawata, 1996; Nakamura *et al.*, 2000), because it can be easily applied to actual three-dimensional problems. Although there have been many simulations focusing on the electric field distribution in the vicinity of the aperture to examine the spatial resolution of NSOM, no calculations have been reported dealing with the light propagation in the tapered region in terms of the sensitivity of the probe. In this paper, using the three-dimensional FDTD method, we demonstrate the high collection efficiency of double-tapered probes including guiding optical fibers, as compared with single-tapered probes. We also examined the dependence of the collection efficiency on the cone angle and taper length in detail.

## 2. Model and Calculations

Figure 1 illustrates the cross-sectional view of the FDTD geometry of the three-dimensional problem, which reproduces the experimental situation of single quantum-dot imaging (Saiki & Matsuda, 1999). A fiber probe with a double- or single-tapered structure collects luminescence ($\lambda=1$ μm) from a quantum dot buried $\lambda/40$ beneath the semiconductor (GaAs; n=3.5) surface. We assume that the source of luminescence is a point-like dipole current linearly polarized along the x direction. The radiation caught by the aperture with a diameter of $\lambda/5$ propagates in the tapered region clad with a perfectly conducting metal and then is guided to the optical fiber waveguide. The refractive indices of the core and cladding of the fiber are 1.487 and 1.450, respectively. The intensity of the collected signal, $I_{coll}$ is evaluated by two-dimensionally integrating the electric field intensity in the core area of the optical fiber. The simulation box consists of a 120x120x360 grid in the x, y, and z directions; the space increment is $\lambda/40$. We run the simulation with a time step of $c\Delta t= \lambda /(40\sqrt{3})$ employing Mur's boundary condition.

## 3. Results and Discussions

To demonstrate the performance of the double-tapered probe, calculations were performed for three types of probes as shown in Fig. 2, where the spatial distribu-tion of the electric field intensity is shown on a logarithmic scale. In Fig. 2(a) and 2(b), $I_{coll}$ is compared for probes with θ=28˚ and θ=90˚. The $I_{coll}$ ratio is estimated to be 1:32. Such a distinct improvement in $I_{coll}$ can be attributed to the difference in the length of the cutoff region. By making the cone angle large and shortening the cutoff region, much radiation power can be directed towards the tapered region. Figure 2(c) shows the result of calculation in the case of double-tapered probe whose cone angle is the same as in Fig. 2(b). The neck diameter D is assumed to be λ, which is twice the cutoff diameter ($d_c$~ λ/2) of the cylindrical waveguide clad with a perfectly conducting metal. $I_{coll}$ of Fig. 2(c) is found to be three times greater than $I_{coll}$ in Fig. 2(b). The radiation pattern in Fig. 2(c) clearly illustrates that the second tapered region modifies the wavefront of the propagating light to match the guiding mode of the optical fiber, while the spherical wave-like propagation in Fig. 2(b) cannot be coupled to the guiding mode so efficiently. To summarize, the collection efficiency of the double-tapered probe in Fig. 2(c) is greater by two orders than that of the conventional single-tapered probe in Fig. 2(a).

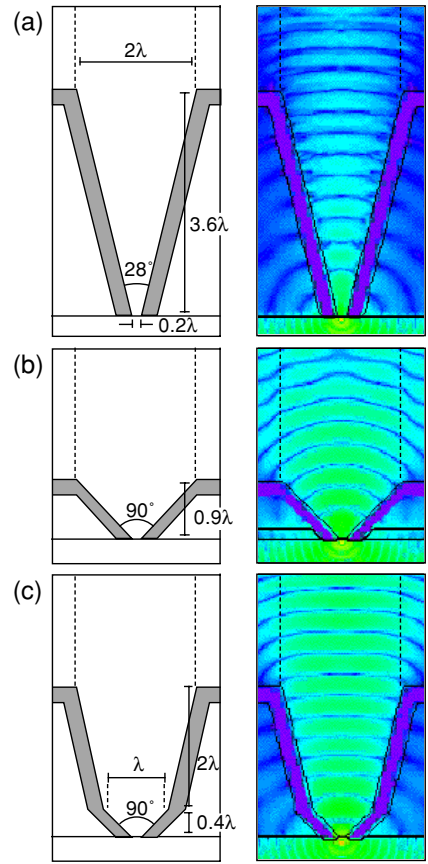

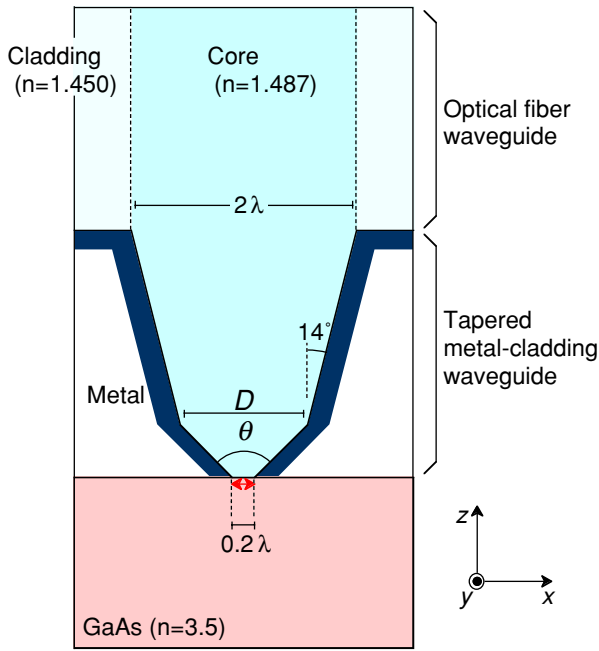

**Figure 1**
Cross-sectional view of the FDTD geometry of the three-dimensional NSOM model.

**Figure 2**
Calculated distribution of the electric field intensity for three types of probes. (a) single-tapered probe with a cone angle θ=28˚, (b) single-tapered probe with θ=90˚, and (c) double-tapered probe with θ=90° and neck diameter D=λ.

Although we have demonstrated the advantage of a double-tapered probe, its performance should be dependent on various structure parameters. In Figs. 3(a) and 3(b), the values of $I_{coll}$ as a function of cone angle θ and neck diameter D, respectively, are plotted The enhancement of $I_{coll}$ with the increase in θ can be understood reasonably. $I_{coll}$ will increase monotonously as θ approaches 180˚. In the case of a realistic metal aperture, however, a large θ will cause diminished spatial resolution due to the finite skin depth of the metal. The optimum value of θ should be chosen by balancing the collection efficiency with the spatial resolution. As depicted in Fig. 3(b), the dependence of $I_{coll}$ on D is found to be more complicated and seems to be less essential. One significant result is that a neck diameter D as small as $d_c$ is more preferable, compared with D~2 $d_c$, to attain high efficiency in coupling to the guiding mode of the optical fiber.

### 4. Summary

FDTD simulation demonstrated the performance of a double-tapered probe, whose collection efficiency was found to be greater by two orders than that of a common single-tapered probe. Such high efficiency could be explained as follows: (1) by shortening the cutoff region of the metal-cladding waveguide, much radiation was directed into the probe; (2) by introducing a guiding region, smooth coupling to the optical fiber was achieved. We also examined the collection efficiency as a function of structure parameters. Dependence on the cone angle was evident as expected; the efficiency increased monotonously with the cone angle. On the other hand, the relationship between the efficiency and the neck diameter was found to be complicated. Further study, focusing on a more realistic situation, introducing a complex dielectric constant of the cladding metal, is now in progress.

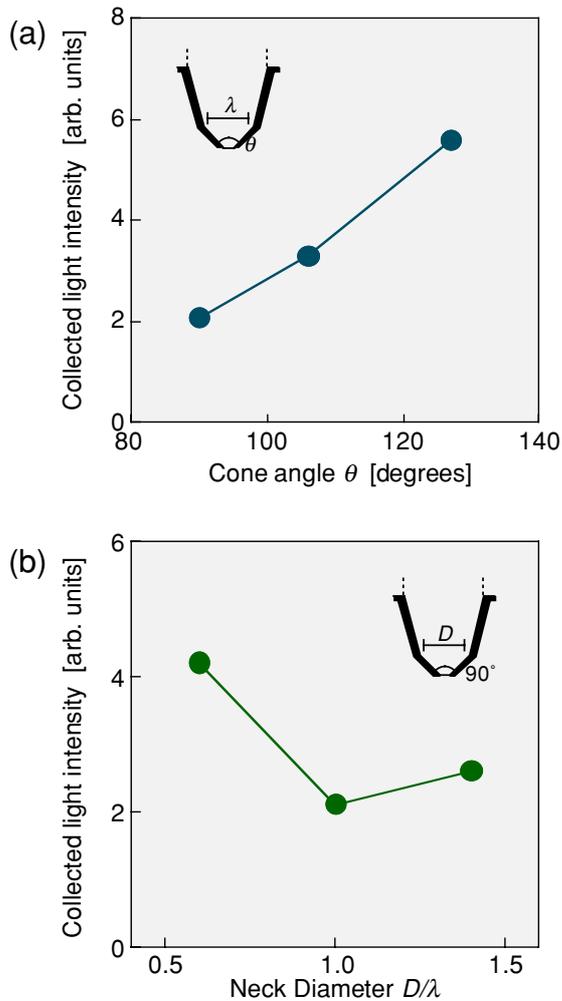

**Figure 3**
Plots of the intensity of collected light as a function of (a) cone angle θ and (b) neck diameter D.

**Acknowledgement**